\documentclass{IEEEtran}
\usepackage{graphicx}  
\usepackage{dcolumn}   
\usepackage{bm}        
\usepackage{verbatim}   
\usepackage{graphicx}
\usepackage{epstopdf}
\usepackage{array}
\usepackage{cite}
\usepackage{psfrag}
\usepackage{booktabs}
\usepackage{footnote}
\usepackage[most]{tcolorbox}
\usepackage{epsfig}
\usepackage{subfigure}
\usepackage{psfrag}
\usepackage{amssymb}
\usepackage{amsmath,bm}
\usepackage{lipsum}
\usepackage{mathtools}
\usepackage{cuted}
\usepackage[utf8]{inputenc}
\usepackage{lipsum} 

\newcolumntype{P}[1]{>{\centering\arraybackslash}p{#1}}

\newcommand{\ve}[1]{\boldsymbol{#1}}

\usepackage[framemethod=TikZ]{mdframed}
\usepackage{lipsum}
	
\begin{document}	
\title{Intelligent-Metasurface-Assisted Full-Duplex Wireless Communications}
\author{Sajjad Taravati and George V. Eleftheriades
	\thanks{}
		\thanks{Sajjad Taravati and George V. Eleftheriades are with the Edward S. Rogers Sr. Department of Electrical and Computer Engineering, University of Toronto, Toronto, Ontario M5S 3H7, Canada.
			Email: sajjad.taravati@utoronto.ca.
		}}

\markboth{}%
{*** \MakeLowercase{\textit{et al.}}: Bare Demo of IEEEtran.cls for IEEE Journals}

\maketitle	
\begin{abstract}
The limited radio-frequency spectrum is a fundamental factor in the design of wireless communication systems. The ever increasing number of wireless devices and systems has led to a crowded spectrum and increased the demand for versatile and multi-functional full-duplex wireless apparatuses. Recently, dynamic and intelligent metasurfaces are explored as a prominent technological solution to the current paradigm of spectrum scarcity by opportunistically sharing the spectrum with various users. In general, intelligent metasurfaces are dynamic, ultra-compact, multi-functional and programmable structures which are capable of both reciprocal and nonreciprocal signal wave transmissions in a full-duplex manner. The controllability and programmability of such metasurfaces are governed through the dc bias and occasionally a radio-frequency (RF) modulation applied to the the active components of the unit cells of the metasurface, e.g., diodes and transistors. This article shows that such intelligent metasurfaces can enhance the performance of wireless communications systems thanks to their unique features such as real-time signal coding, nonreciprocal-beam radiation, nonreciprocal beamsteering amplification, and advanced pattern-coding multiple access communication.
\end{abstract}

\section{Introduction}

Conventional wireless systems operate based on transmission and reception of data in different frequency sub-bands or time slots. However, such a half-duplex communication results in a communication delay and a crowded spectrum. To overcome this issue and enhance the communication efficacy, full-duplex communication systems have been proposed for supporting simultaneous transmission and reception in narrow time and frequency slots~\cite{zhang2016full}. Full-duplex communication is a technique to attain higher spectral efficiency, representing a standalone solution or a complementary technique to orthogonal frequency division multiplexing (OFDM), multiple-input–multiple-output (MIMO) and mm-wave transceivers. Compared to conventional frequency-division duplex systems, which utilize different channels for transmission and reception, full-duplex operation improves the attainable spectral efficiency by a factor of two.

Recently, researchers from various areas such as electromagnetics, communication theory, and circuit and device implementation techniques have reported remarkable progress in enhancing spectral efficacy and exploiting underutilized spectrum to attain higher data rates. Nevertheless, such techniques usually require radio complexity, which in turn implies higher power consumption and cost in comparison with established counterparts. For instance, MIMO transceivers enhance the throughput by leveraging spatial diversity while requiring cumbersome transceiver array elements. On the other hand, millimeter-wave transceivers offer high data rates while suffering from high path loss and unfavorable propagation characteristics. This in turn yields a much more complex system architecture and higher power consumption.

The ever-growing communication traffic is caused by emerging massive-data applications, e.g., autonomous vehicles, tactile Internet, virtual reality, and augmented reality. However, 5G communication systems are incapable of efficiently supporting such applications. Potential technologies for 6G mobile communication systems include terahertz-band communications, ultra-massive MIMO, blockchain-based spectrum sharing, quantum computing, orbital angular momentum multiplexing, and programmable intelligent-metasurface-based signal sharing~\cite{dang2020should,zhou2020joint}. Intelligent metasurfaces are dynamic, reconfigurable, programmable, potentially nonreciprocal and signal-boosting engineered surfaces which are capable of steering electromagnetic waves in a full-duplex manner. They represent an emerging technology for tailored and advanced coding of electromagnetic waves.  

\begin{figure*}
	\begin{center}
		\subfigure[]{\label{fig:conc}
		\includegraphics[width=1.2\columnwidth]{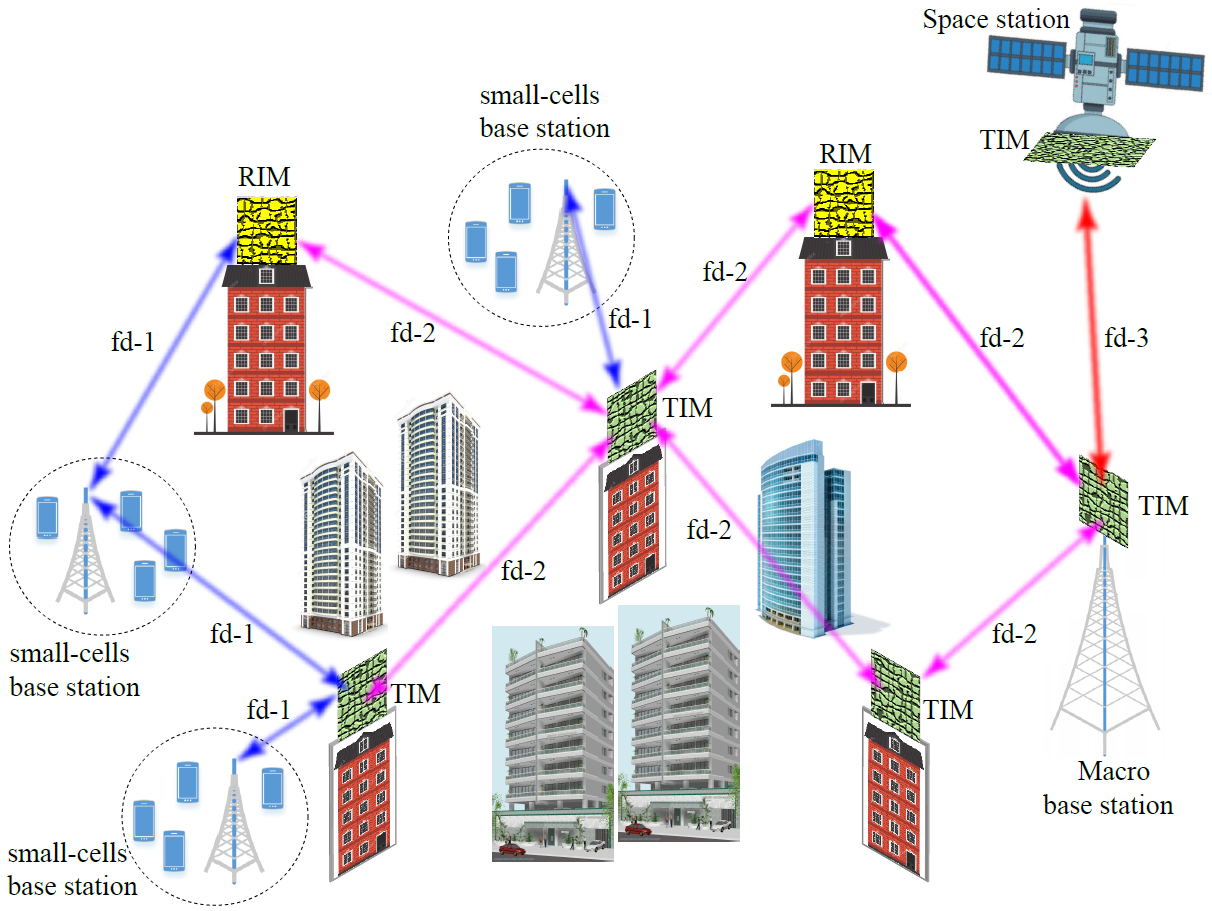}}	
		\subfigure[]{\label{fig:conc_b} 
		\includegraphics[width=1.5\columnwidth]{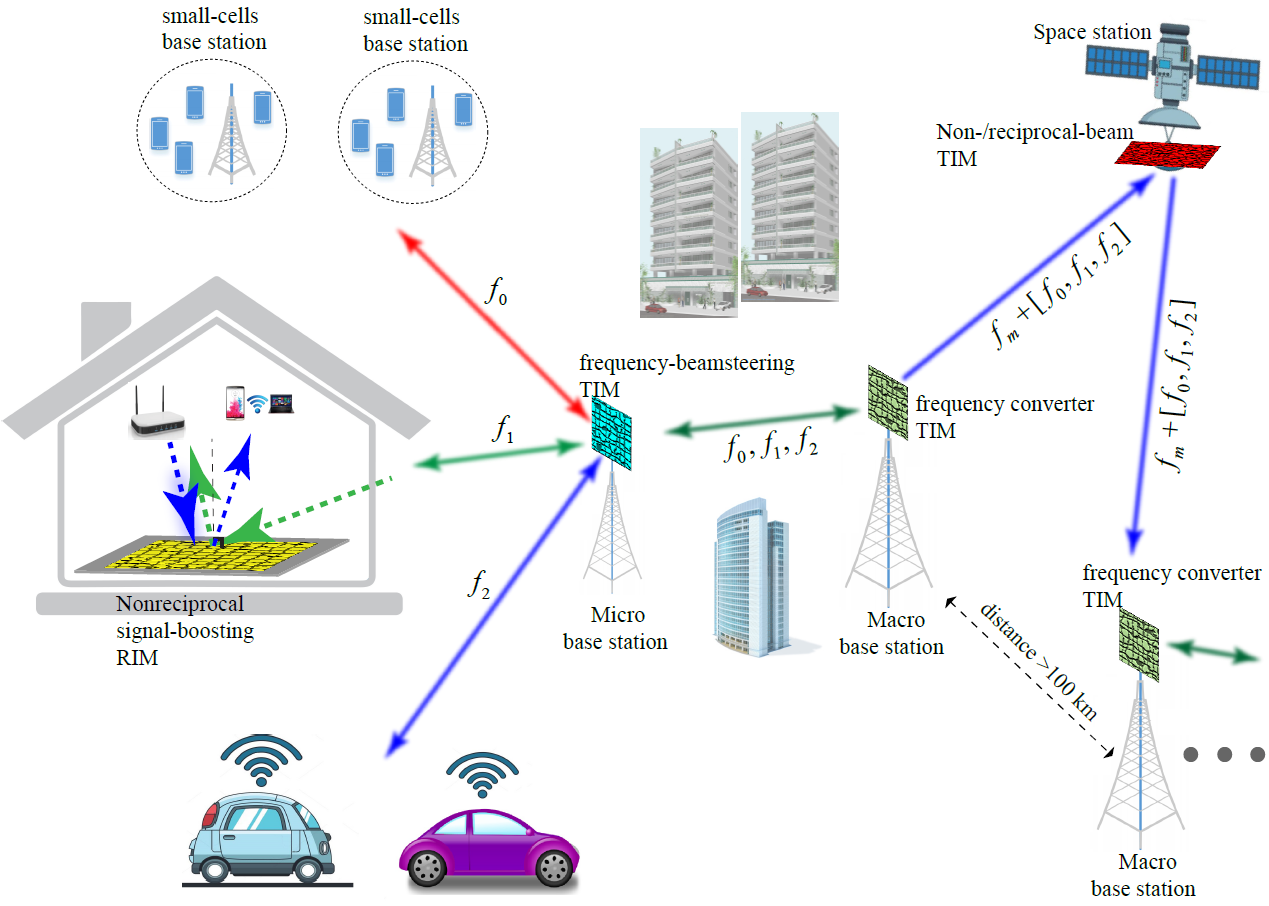}}	
		\caption{Full-duplex wireless communication in urban areas based on transmissive and reflective intelligent metasurfaces (TIMs and RIMs). (a) Generic scenario. (b) Based on specific active multi-functional metasurfaces.}
		\label{fig:concc}
	\end{center}
\end{figure*}

Intelligent dynamic metasurfaces have recently drawn enormous attention due to their extraordinary capability and potential in wave processing across a wide frequency range, from microwaves to terahertz~\cite{taravati_PRApp_2019,ye2020joint,tsilipakos2020toward,taravati2020full,liu2020joint,Taravati_AMA_PRApp_2020,wang2020intelligent,taravati2021full,zhang2021wireless,taravati2021pure,wan2021terahertz}. Such intelligent metasurfaces are composed of subwavelength constituents with controllable parameters and adjustable responses (e.g., phase, amplitude, angle of transmission, polarization and operation frequency) through an external bias-field in a real-time reconfigurable manner. Over the past few years, an ever increasing research effort has been made on the applications of intelligent metasurfaces to wireless communications. Intelligent metasurfaces have been employed for the realization of novel and highly efficient wireless transceiver schemes, resulting in a paradigm shift of the transceiver design and reduced hardware cost of the wireless communication systems. In particular, microwave intelligent-metasurface-based wireless communication systems are of great interest. This is because microwave frequencies represent the most favorable frequency ranges for wireless communication, due to the relatively low path loss, non line-of-sight operation and small size of transceiver components (e.g., antennas and transmission lines). This study presents a comprehensive list of intelligent-metasurface-based communication systems which provide an exceptional platform for efficient and programmable data transmission and coding in a full-duplex manner.

Currently, prototypes of reconfigurable and programmable metasurfaces are being demonstrated and startup companies have embarked developing the fundamental technology that covers a wide range of applications. This paper overviews and analyzes different reconfigurable metasurfaces with distinct features and functionalities, that can cover a wide range of requirements of future intelligent-metasurface-assisted full-duplex communications. We describe different possible applications of such reconfigurable and intelligent metasurfaces enabling full-duplex urban, satellite and cellular, wireless communications.

\section{Intelligent-Metasurface-Based Communication Schemes}

Intelligent metasurfaces may represent a fundamental component in future full-duplex wireless communications. An electromagnetic metasurface is a two-dimensional array made of unit cells that are engineered in a way to introduce properties that are not readily available in naturally occurring materials. Reconfigurable and intelligent metasurfaces are deeply subwavelength in thickness and electrically large in their transverse size, and comprise dynamic scattering unit cells. The specific arrangements and bias of the dynamic unit cells determine how the intelligent metasurface transforms the incident wave into specified but arbitrary reflected and transmitted electromagnetic waves. Such metasurfaces should be capable of providing both reciprocal and nonreciprocal transmission and reflection of signal waves, as well as altering the spectrum of the incident wave by generating specified new frequencies. The unit cells of intelligent metasurfaces may be loaded by active components such as diodes (e.g., varactor, pin and Schottky diodes) or transistors which may be biased through both direct current (dc) and occasionally augmented by RF modulation signal waves. The dc biasing signal is usually employed to change the operation point of the diodes (e.g., average capacitance of varactors) while the RF biasing signal provides a time modulation signal for temporal or spatiotemporal modulation of the effective permittivity or conductivity. Both dc and RF biasing signal waves can then be leveraged for controlling the characteristics and response of the unit cells, and the operation and functionality of the metasurface. Eventually, the real-time full control of the intelligent metasurface can be carried out by a computer processor possibly running an artificial intelligence (AI) application or a field-programmable gate array (FPGA).

Figure~\ref{fig:conc} illustrates a generic scenario for an intelligent-metasurface-assisted full-duplex wireless communication in urban areas. In this scheme, transmissive and reflective intelligent metasurfaces (TIMs and RIMs) are placed on top of specified buildings to facilitate an effective full-duplex communication between different parts of the network, that is, small-cell base stations, macro base stations and satellites. Here, TIMs and RIMs are capable of introducing various functionalities such as intelligent beamsteering, signal wave amplification, frequency conversion, channel allocation and coding. In Figure~\ref{fig:conc}, the full-duplex communications links between small-cell base stations and intelligent metasurfaces are shown with blue double arrows and denoted by fd-1. The full-duplex communications links between two intelligent metasurfaces are shown with magenta double arrows and denoted by fd-2, and the full-duplex communications links between macro base station and satellite TIM are shown with red double arrows and denoted by fd-3. RIM and TIM in each of these three full-duplex links may possess distinct characteristics and functionalities such as beamsteering, signal boosting, frequency conversion, real-time signal coding, multiple access communication and nonreciprocal transmission and reflection. The required functionality by each metasurface is governed by the core of the system in a software-controlled and programmable manner, e.g., via AI. 


An efficient and intelligent real-time beamsteering guarantees that all the static and mobile elements of the communication system are efficiently covered, whereas signal boosting and amplification enhances the efficiency and quality of the communication. Furthermore, frequency conversion may be required in some parts of the communication network for the sake of bandwidth enhancement or adaptation of the signal wave to the communication channel for lower path loss. Real-time signal coding and multiple access communication are also essential parts of a secure communication network that may be required to guarantee a secure link. Finally, nonreciprocal signal wave transmission and nonreciprocal-beam radiation are appealing functionalities that drastically increase the flexibility and capacity of the communication network. Figure~\ref{fig:conc_b} illustrates a possible scenario of an intelligent-metasurface-assisted full-duplex wireless communication in urban areas using active multi-functional metasurfaces. In this scenario, nonreciprocal, frequency converting and frequency-beamsteering intelligent metasurfaces are employed to enhance the efficiency of the communication network. In the following sections, we elaborate on some of the reconfigurable TIMs and RIMs that are capable of providing the aforementioned functionalities.

\section{Beamsteering and Signal Boosting Nonreciprocal RIM}
\begin{figure}
	\begin{center}
		\includegraphics[width=1\columnwidth]{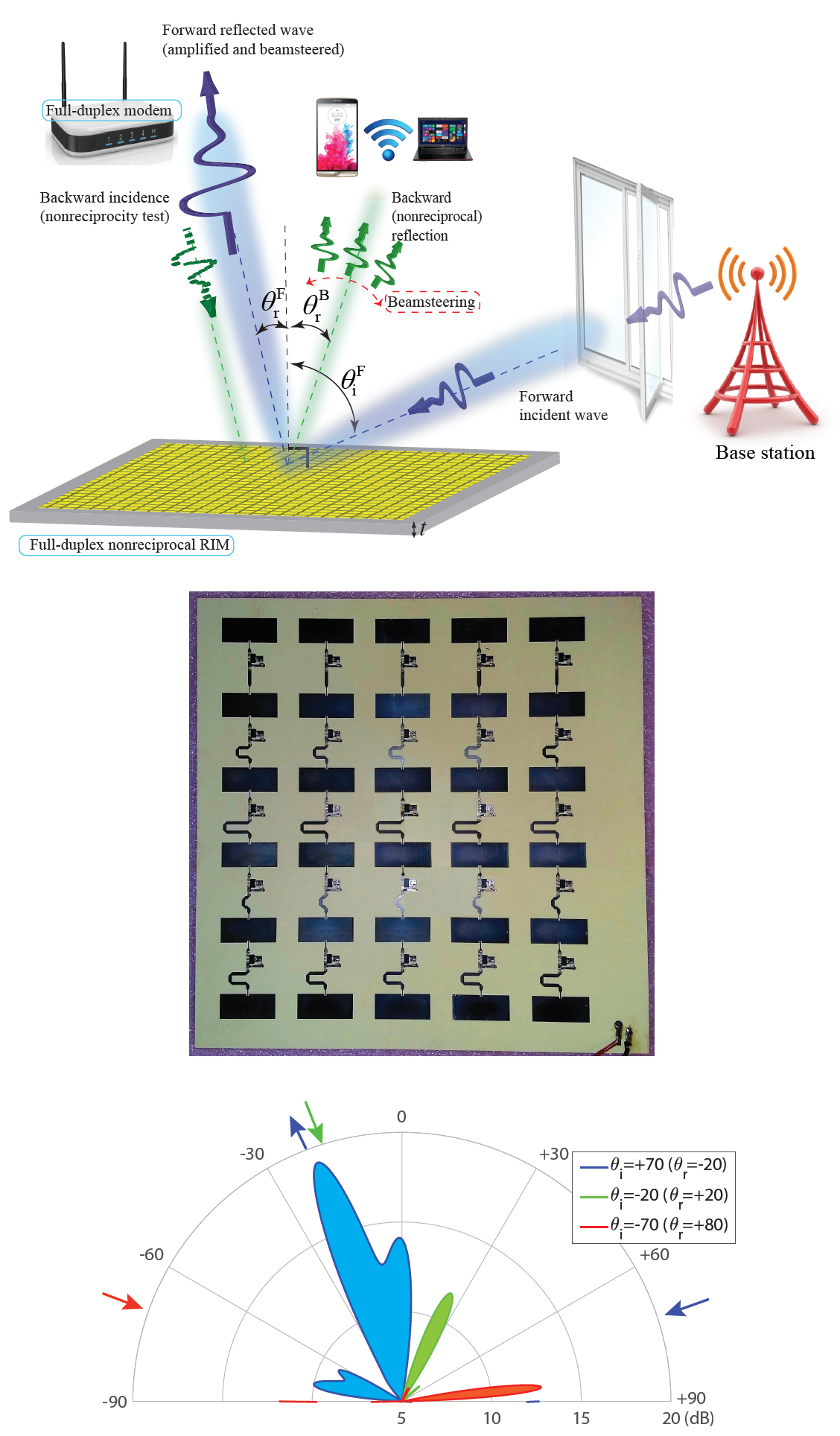}	
		\caption{Reflective intelligent metasurface for full-duplex nonreciprocal beamsteering and amplification. (top) RIM functionality, and (bottom) fabricated prototype and measured results.}
		\label{fig:MeasSU}
	\end{center}
\end{figure}

Figure~\ref{fig:MeasSU} shows a reflective intelligent metasurface (RIM) for full-duplex nonreciprocal beamsteering and amplification~\cite{taravati2021full}. This RIM is composed of an array of phase-gradient cascaded radiator-amplifier-phaser unit cells. The unit-cell of this RIM is composed of a patch radiator, a unilateral circuit and a reciprocal phase shifter. An electromagnetic wave impinging on top of the RIM from the right side (forward incident wave) under the angle of incidence $\theta_\text{i}^\text{F}$ is reflected and steered towards a desired direction in space under the angle of reflection $\theta_\text{r}^\text{F}$, while possessing the same frequency as the frequency of the impinged wave. The backward incidence assumes the time-reversed version of the forward incidence, where an electromagnetic wave impinges on top of the RIM from the left side under the angle of incidence $\theta_\text{r}^\text{F}$. Then, it is amplified by the RIM and is being reflected towards a desired direction in space under the angle of reflection $\theta_\text{r}^\text{B} \neq \theta_\text{i}^\text{F}$. Such a nonreciprocal reflection is evidently different than the reflection by a reciprocal surface where $\theta_\text{r}^\text{B} = \theta_\text{i}^\text{F}$. The metasurface is composed of a dielectric layer sandwiched between two conductor layers, where the conductor layers are constituted of a plurality of unit cells embedded therein. The operation of the unilateral transistors and unit cells is controlled via a DC biasing signal, thereby arbitrary phase shift and amplitude profiles can be created along the metasurface.

Such RIMs can be mounted on a roof, wall or on a smart device in a seamless manner. They may be further developed for massive MIMO beam-forming, where multiple beams and multiple frequencies are supported, and no cumbersome RF feed lines and matching circuits are required. The RIM functionality is fully controlled through dc biasing of the transistors, variable phase shifters, and tunable patch radiators. Such a highly directive and reflective full-duplex nonreciprocal-beamsteering represents a promising feature to be utilized for low-cost high capability and programmable wireless beamsteering applications. Such an RIM may act as the core of an intelligent connectivity solution for signal boosting in cellular, WiFi and satellite receivers and Internet of Things (IoT) sensors, introducing high speed scanning between users, full-duplex multiple access and signal coding.

\section{Nonreciprocal-Beamsteering TIM}

\begin{figure}
	\begin{center}
		\includegraphics[width=1\columnwidth]{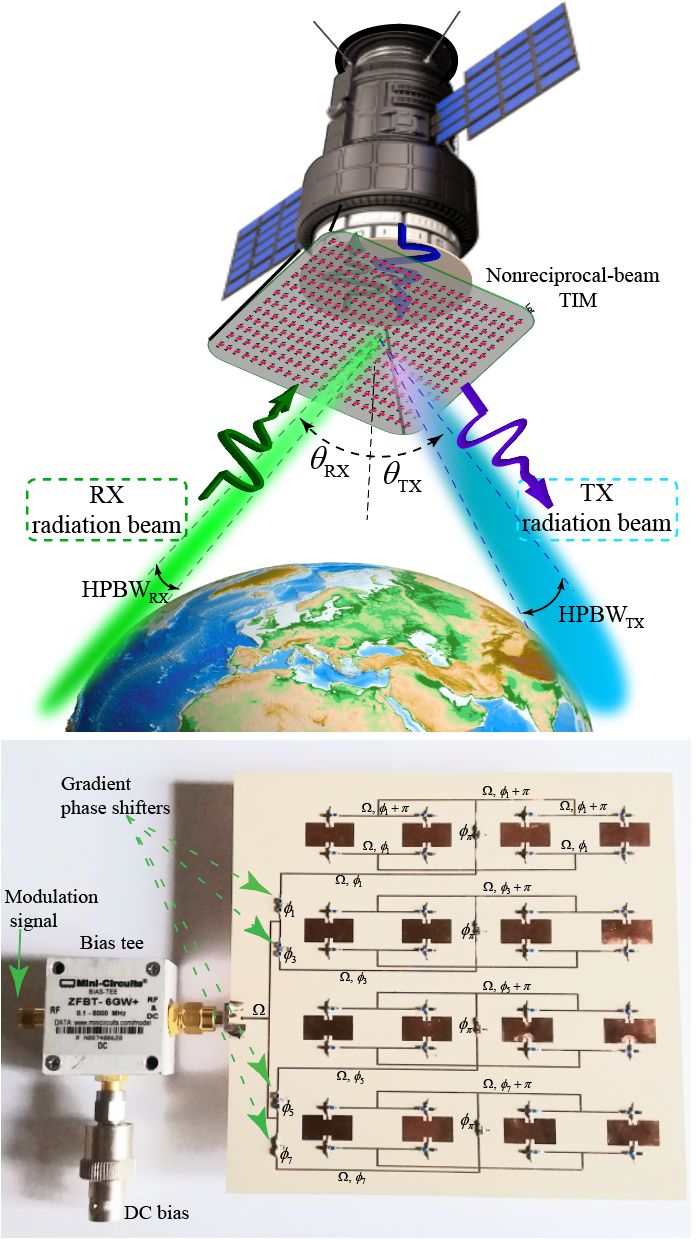}	
		\caption{Full-duplex nonreciprocal-beam TIM for efficient point-to-point wireless communications.}
		\label{Fig:NRBM}
	\end{center}
\end{figure}

Figure~\ref{Fig:NRBM} shows a nonreciprocal-beam TIM for efficient full-duplex point-to-point wireless communications~\cite{taravati2020full}. Such a TIM provides different radiation beams for transmission and reception, i.e., different radiation angles and half-power-beam-widths (HPBW), as $F_\text{TX} (\theta)\ne F_\text{RX}(\theta)$, where $F_\text{TX} (\theta)= E_{\theta,\text{TX}} /E_{\theta,\text{TX}}(\text{max})$ is the transmission radiation pattern, and $F_\text{RX}(\theta)=E_{\theta,\text{RX}}/E_{\theta,\text{RX}}(\text{max})$ is the reception radiation pattern. The realization of such a full-duplex nonreciprocal TIM is accomplished through an array of time-modulated phase-gradient unit cells. 

Such a functionality is achieved by leveraging the unique properties of asymmetric frequency-phase transitions in time-modulated unit cells, where all the unwanted time harmonics are suppressed in an elegant manner. In addition, there is no inherent  limit to the bandwidth enhancement of the proposed TIM. The time modulation is practically achieved through biaisng of varactor diodes by a dc signal and an RF time-varying signal. The dc and RF signals are then used for controlling the functionality of the metasurface. For instance, the transmission and reception beams of the metasurface can be steered through the phase shift of the RF modulation signal bias, and the magnitude of the transmission and reception beams can be tuned through the amplitude of the RF modulation signal bias.

\section{Frequency-Beamsteering TIM}

\begin{figure}
	\begin{center}
		\subfigure[]{\label{fig:pr1} 
			\includegraphics[width=1\columnwidth]{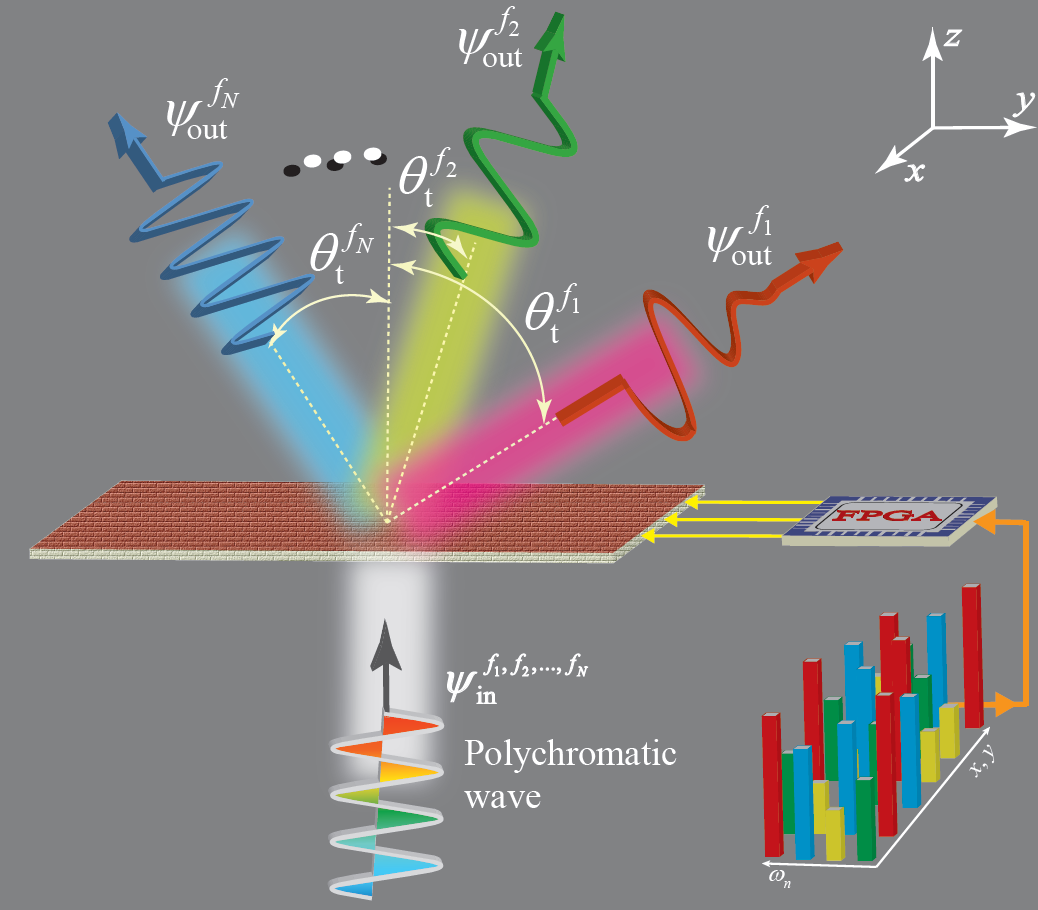}} 
		\subfigure[]{\label{fig:pr2} 
			\includegraphics[width=1\columnwidth]{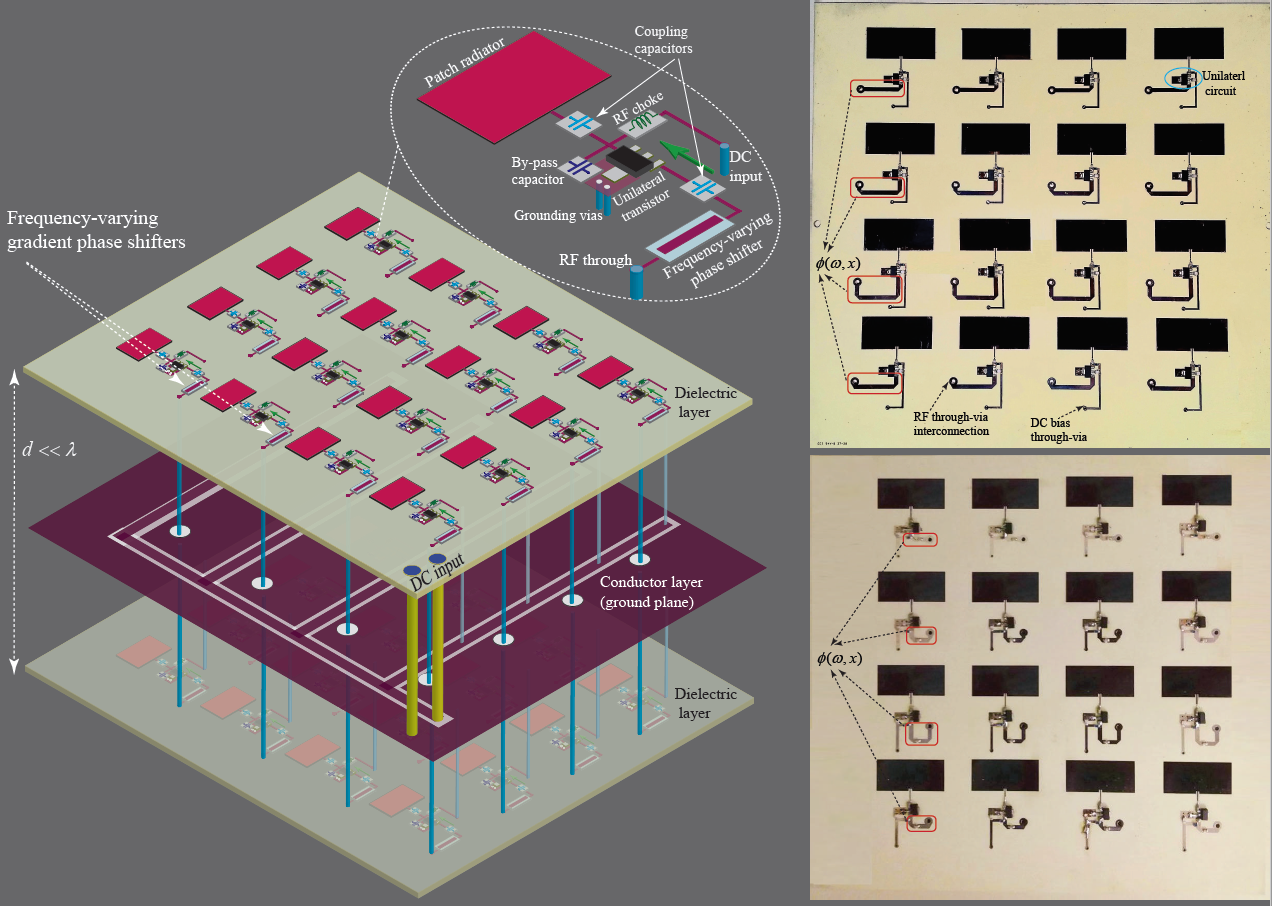}} 
		\caption{Frequency-beamsteering TIM. (a) Functionality. (b) Fabricated prototype.}
		\label{fig:prism}
	\end{center}
\end{figure}

Figure~\ref{fig:pr1} sketches the functionality of a frequency-beamsteering TIM. Such a TIM provides a frequency-dependent spatially variant phase shift and a frequency-dependent spatially variant signal gain~\cite{taravati2021programmable}. We assume a polychromatic electromagnetic wave $\ve{\psi}_\text{in}$ that comprises various frequency components, $\omega_1,\omega_2,...,\omega_\text{N}$, passes through the TIM, where the TIM introduces frequency-dependent spatially variant phase shifts. Hence, the frequency components of the incident polychromatic wave acquire different phase shifts at each point on the $x-y$ plane, i.e., $\phi_n(x,y)$. The phase and magnitude profiles of the TIM, i.e., $\phi(\omega,x,y)$ and $T(\omega,x,y)$, are engineered is a way that each transmitted frequency component $\ve{\psi}_\text{out}^{f_n}$ is transmitted under a desired transmission angle $\theta_\text{t}^{f_n}$ with a specified signal gain of $|\ve{\psi}_{\text{out},n}|$.

Figure~\ref{fig:pr2} shows the architecture and a photo of the fabricated frequency-beamsteering TIM. The frequency-beamsteering TIM is composed of an array of active frequency-dependent spatially variant unit cells introducing the required  phase shift and magnitude for efficient beamsteering and signal amplification. The experimental results show favorable features such as prism-like spatial decomposition of frequency components, possible isolation between the forward and backward transmissions, and a desirable signal gain. Furthermore, the frequency-beamsteering TIM is fairly broadband and its bandwidth can be further increased by standard band-broadening methods.

\section{Diffraction Code Multiple Access TIM-Assisted System}\label{sec:app}

\begin{figure}
	\begin{center}
		\includegraphics[width=1\columnwidth]{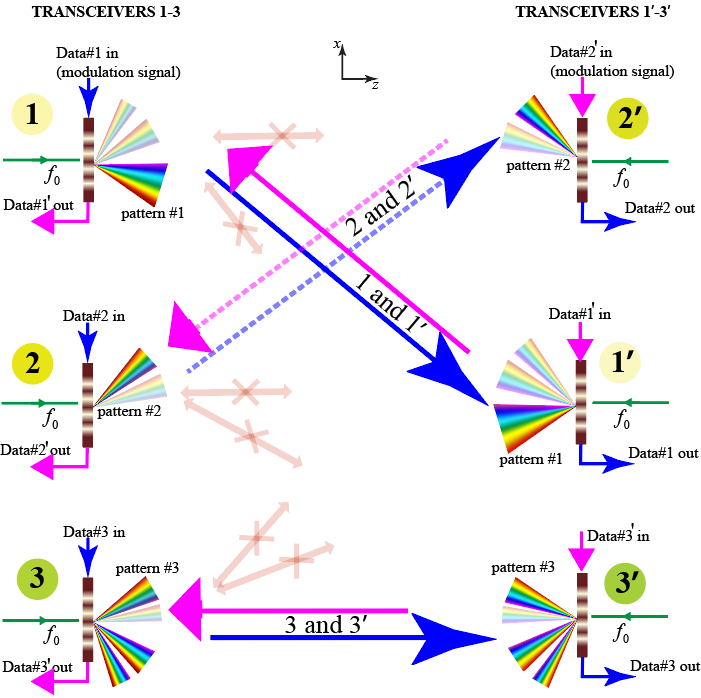}
	\caption{Space-time diffraction-code multiple access system based on real-time pattern coding of space-time diffraction TIMs.}
	\label{fig:coding}
		\end{center}
\end{figure}

Figure~\ref{fig:coding} sketches a full-duplex space-time diffraction-code multiple access system which is based on real-time diffraction pattern generation of space-time diffraction TIMs~\cite{taravati_PRApp_2019}. Such a coding TIM-assisted system offers unique properties that can be utilized for advanced and versatile real-time signal coding and beamsteering. In the particular example in Fig.~\ref{fig:coding}, three pairs of transceivers are considered, but in practice one may consider more pairs of transceivers. Then, only the transceiver pairs sharing identical space-time diffraction patterns are communicating effectively. Each diffraction pattern is attributed to the properties of the space-time-modulated TIM, that is, the input frequency, where the input data (message) plays the role of the modulation signal. For a specified input modulation data signal, a unique diffraction pattern is achieved. Here, the transceiver pairs that are allowed to communicate are $1$ and $1'$, $2$ and $2'$, and $3$ and $3'$, whereas the transceivers $2'$ and $3'$ ($2$ and $3$) cannot retrieve the data sent by the transceiver $1$ ($1'$), and the transceivers $1'$ and $3'$ ($1$ and $3$) cannot retrieve the data sent by the transceiver $2$ ($2'$). Distinct diffraction patterns can be created by certain space-time modulation parameters, e.g., the modulation amplitude, modulation frequency and TIM thickness. In addition, the radiation pattern provided by a space-time TIM is very diverse and is sensitive to the space-time modulation parameters. Hence, an optimal isolation between the transceivers can be achieved by proper engineering of the space-time diffraction patterns.

\section{Frequency Converter Metasurface}	

\begin{figure}
	\begin{center}
			\includegraphics[width=1\columnwidth]{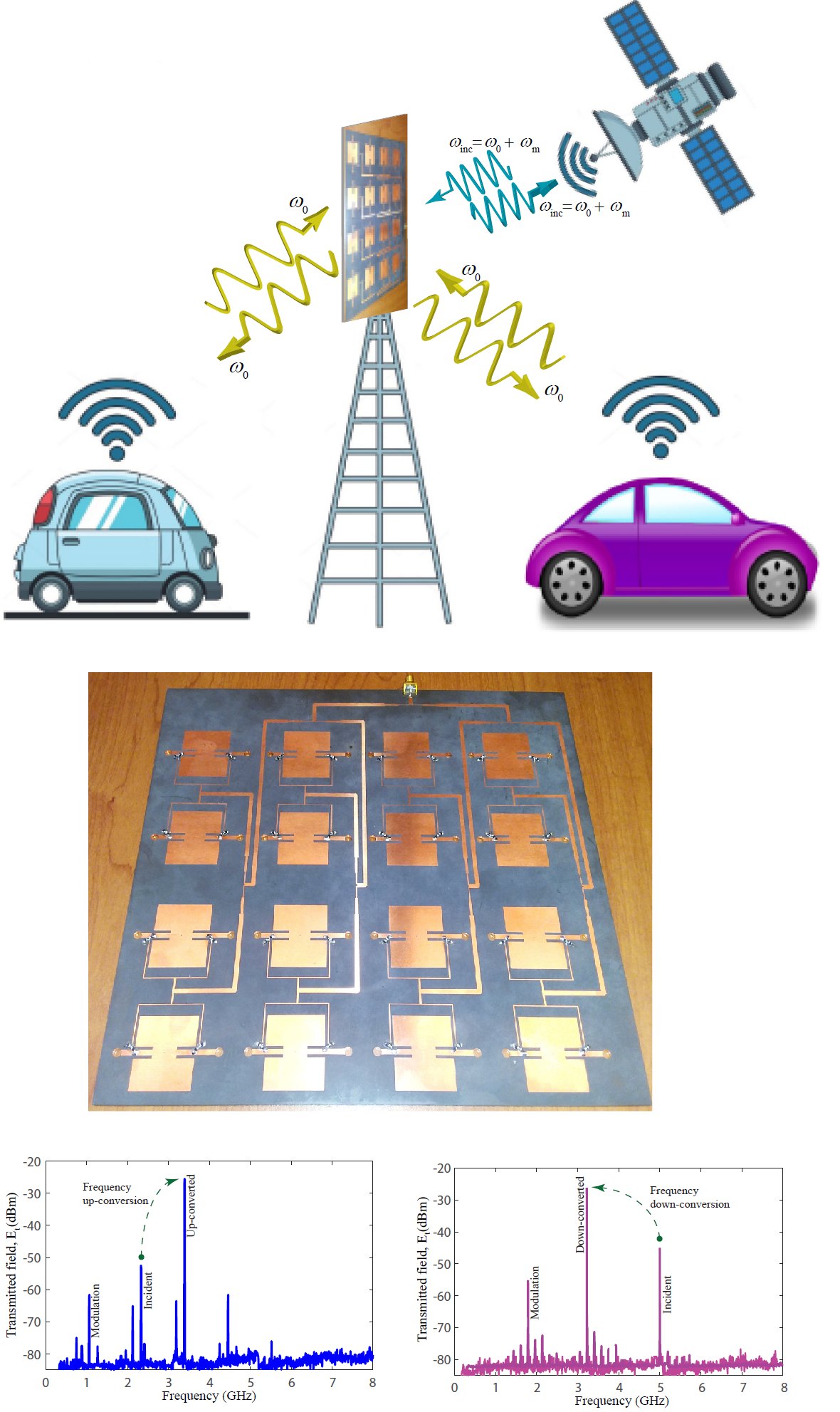}
		\caption{Experimental results of a pure frequency converter temporal TIM for full-duplex wireless communications.}
		\label{fig:FC}
	\end{center}
\end{figure}

Frequency conversion is a crucial task in wireless communications for adaptation of the transmitted signal depending on the communication channel characteristics for lower path loss. Figure~\ref{fig:FC} shows the application of a frequency converter TIM to satellite and cellular communications~\cite{taravati2021pure}. Here, the metasurface is placed on top of a base station to create an efficient connection between the cellular base stations and mobile phones and satellites. In particular, consider the cellular phones operating in Broadband Personal Communications Service (PCS Band), i.e., 1850-1910/1930-1990 MHz and the satellite network operating at C-band, i.e., 3700 and 7025 MHz which is a well-known frequency band for satellite communications. Hence, the TIM makes an efficient connection between the cellular and satellite networks. To improve the functionality of the temporal TIM in this application, appropriate beamsteering mechanisms~\cite{taravati2020full,taravati2021full,taravati2021programmable} can be added to the frequency conversion functionality of the TIM.

\section{Concluding Remarks}

Reconfigurable intelligent metasurfaces are energy-efficient structures that are capable of reshaping the wireless environment for an enhanced throughput. They introduce appealing features and functionalities such as gain, beamsteering, spatial frequency decomposition, nonreciprocal signal transmission, frequency conversion, nonreciprocal-beam radiation, and real-time pattern coding. Such features make reconfigurable TIM and RIM a promising solution for enabling future communication-intensive applications such as smart cities. These structures can optimize the spectral efficiency for single and multiple carrier communication systems. In addition, they can provide a complex bit/power allocation on different spatial-domain sub-channels for an improved bit error rate performance, and present a spatial-decomposition-based beamforming and beamsteering platform for MIMO systems. The main restriction of current research on reconfigurable-intelligent-metasurface-based wireless communications is the lack of realistic prototypes and experiments on practical multifunctional RIMs and TIMs. In this article, we have discussed some recent experimental results on different reconfigurable intelligent metasurfaces that can be utilized for a variety of applications in 5G/6G wireless communications and beyond.

\bibliographystyle{IEEEtran}
\bibliography{Taravati_Reference}

\vspace{5mm}
	\begin{center}
\textbf{BIOGRAPHIES}
\end{center}

\noindent Sajjad Taravati [SM'21] (sajjad.taravati@utoronto.ca) received his Ph.D. degree in electrical engineering from Polytechnique Montreal, Montreal, QC, Canada. He is currently a Post-Doctoral Fellow at the University of Toronto, Toronto, Ontario, Canada. His current research interests include all fields of telecommunications, electromagnetic structures, antennas, nonreciprocal magnetless systems, space–time-modulated structures, electronic circuits and systems, broadband phase shifters, wave engineering, airborne telecommunication transceivers, microwave active and passive circuits, power amplifiers, radio modules, electromagnetic theory, invisibility cloaking, and active metasurfaces.

\vspace{3mm}

\noindent George V. Eleftheriades [F'06] (gelefth@waves.utoronto.ca) earned the Ph.D. degree in
electrical engineering from the University of Michigan, Ann Arbor, MI, USA, in 1993. Currently, he is a Professor in the
Department of Electrical and Computer Engineering at the University of Toronto, ON, Canada. His research has impacted the field by demonstrating the unique electromagnetic properties of metamaterials; used in lenses, antennas, and other
microwave and optical components to drive innovation in fields such as satellite communications, defence, medical imaging, microscopy, automotive radar, and wireless telecommunications.

\end{document}